# Change of quantum correlation for two simultaneously accelerated observers

Yue Li , Yongjie Pan , and Baocheng Zhang*

School of Mathematics and Physics, China University of Geosciences, Wuhan 430074, China

*Email: zhangbc.zhang@yahoo.com

**Abstract.** The influence of Unruh effect on the quantum and classical correlation of a quantum entangled state is investigated, when one or two of the observers are accelerated. It is found that the quantum and classical correlation would approach to a finite value with the increase of the acceleration when only one observer is accelerated, but they would decrease to zero quickly when two observers are accelerated. The latter result implies that the gravitational field would break the correlation obviously when the acceleration is caused by the gravitational field. Thus, when the two observers stay in the gravitational field, the quantum or classical correlation would not easy to be preserved for a long time.

**1. Introduction**

Quantum mechanics and general relativity are the two pillars underpinning modern physics. Although they have solved many important problems in physics, the contradiction between these two theories is also quite striking. In order to solve this problem, it is particularly important to study the quantum effects in the context of curved space-time. In 1976, Unruh proposed the Unruh effect [1] in his research on quantum field theory. This effect states that an observer under the influence of the acceleration in the Minkowski space-time would associate a thermal bath to the vacuum state of the inertial observers. The accelerated observers will feel a Unruh temperature,

$$k_B T = \frac{\hbar a}{2\pi c} \quad (1)$$

where $k_B$ is the Boltzmann's constant, $\hbar$ is the reduced Planck constant, $c$ is the speed of light in a vacuum, and $a$ is the acceleration.

Since the Unruh effect was put forward, it has been studied in many different aspects (see the review [2] and references therein). Recently, the effect of Unruh effect on quantum correlation and entanglement has caused much attention [3-13], due to the rapid development of quantum information theory. It is noted that the change of quantum entanglement and the quantum correlation quantified by the quantum discord when one of two observers is accelerated [16] and the quantum entanglement when two observers are accelerated simultaneously [9] had been investigated in detail, but the quantum and classical correlation has not been studied for two simultaneously accelerated observers. In this paper, we will investigate this. Based on a recent suggestion that the Unruh effect can be regarded as a noise-induced quantum channel by a complete positive super-operator, we would further analyze and study this channel from the perspective of correlation. It is found interestingly that the classical and quantum correlation would decrease into zero quickly with the increase of the acceleration for the case that two observers accelerates simultaneously, but the result is not like that for the case that only one observer is accelerated.

The structure of the paper is arranged as follows. In the second section, we review the earlier views on the Unruh effect as a quantum noise channel but with the different initial state. In the third section, we introduce the description about the quantum and classical correlation for the general quantum mixed state, and calculate the mutual information, quantum discord [14], and the classical correlation [15] for the quantum state influenced by the acceleration of one observer. In the fourth section, we describe the evolution of the system by allowing not only Rob but also Alice to experience uniform acceleration. The same study will be carried out using the same method of studying in the third section. And then the two cases were compared. At the end of the paper, it is summarized and extended.

**2. Unruh effect as quantum noise channel**
Recently, the Unruh effect was considered as a quantum noise channel, and the equivalent map and the corresponding entanglement fidelity was discussed [16]. In this section, we will review their work but with the different quantum initial state,

$$|\psi\rangle = \frac{1}{\sqrt{2}}\left(|0\rangle_A |0\rangle_R + |1\rangle_A |1\rangle_R\right) \tag{2}$$

where the subscripts $A$ and $R$ represents the Minkowski modes seen by the observers Alice and Rob. When Rob begins to accelerate, the state (2) will be changed according to the following forms,

$$|0\rangle_R = \frac{1}{\cosh r}\sum_{n=0}^{\infty}\tanh^n r |n\rangle_I \otimes |n\rangle_{II}$$

$$|1\rangle_R = \frac{\sqrt{n+1}}{\cosh^2 r}\sum_{n=0}^{\infty}\tanh^n r |(n+1)\rangle_I \otimes |n\rangle_{II} \tag{3}$$

where $r$ is the acceleration parameter with its definition, $\tanh r = e^{-\pi\frac{kc}{a}}$, $|n\rangle_I$ and $|n\rangle_{II}$ refer to the modes in the Rindler wedge I and wedge II [20-22], and $|1\rangle_R$ is an excitation of the Minkowski mode $|0\rangle_R$. Then, the quantum state (2) after acceleration becomes

$$\rho' = \sum_{n=0}^{\infty} C_n \left[|0n\rangle\langle 0n| + \frac{\sqrt{n+1}}{\cosh r}\left(|0n\rangle\langle 1(n+1)| + |1(n+1)\rangle\langle 0n|\right) + \frac{n+1}{\cosh^2 r}|1(n+1)\rangle\langle 1(n+1)|\right]$$

(4)

where $C_n = \frac{\tanh^{2n} r}{2\cosh^2 r}$, the subscripts $A$ and $R$ has been omitted for brief, and through the paper, when the subscripts are omitted, the former one is the mode seen by Alice and the latter one is the mode seen by Rob in the form $|1n\rangle$ for example.

As in the Ref.[16], the change of state by the acceleration can be regarded as a similar mechanism to the quantum noise channel described by a complete positive super-operator

$$A_n = \frac{1}{\sqrt{n!}}\tanh^n r \left(\cosh r\right)^{-(\hat{n}_A + 1)} \otimes \left(b^\dagger\right)^n \tag{5}$$

where $\hat{n}_A = a_A^\dagger a_A$ is a number operator acting on Alice's Hilbert space and $b$, $b^\dagger$ is the creation, annihilation operators on Rob's Hilbert space but in the background of Rindler spacetime. The relation is given by the Bogoliubov transformation [21].

Thus, the Unruh effect can be considered as a noise quantum channel

$$\rho \to \rho' = \sum_n A_n \rho A_n^\dagger$$

where $\rho = |\psi\rangle\langle\psi|$. It is easy to prove that the operator $A_n$ is trace preserving as in Ref. [16]. The entanglement fidelity of the quantum noise channel is calculated as,

$$Fe = \sum_n (Tr\rho A_n)(Tr\rho A_n^\dagger) = \frac{1}{4} \frac{1}{\cosh^2 r} \left(1 + \frac{1}{\cosh r}\right)^2$$

Although the expression is different from that in Ref. [16], its change trend with the acceleration is the same, as seen from the black dashed line in Fig.1.

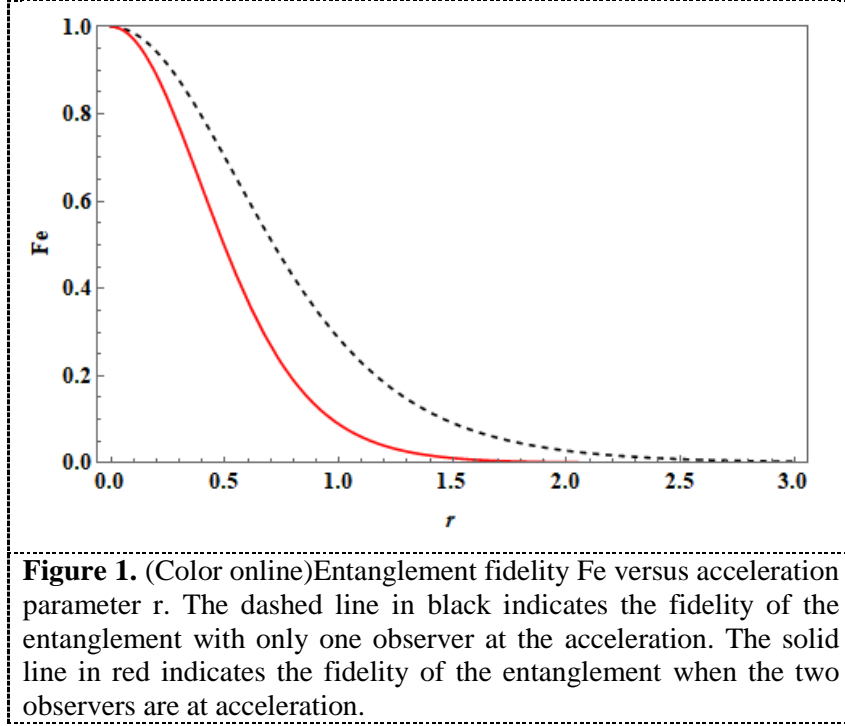

**Figure 1.** (Color online) Entanglement fidelity Fe versus acceleration parameter r. The dashed line in black indicates the fidelity of the entanglement with only one observer at the acceleration. The solid line in red indicates the fidelity of the entanglement when the two observers are at acceleration.

## 3. Quantum and Classical correlation

The variation of quantum state correlation caused by Unruh effect can be described by mutual information. Mutual information is a physical quantity measuring correlation, which includes both quantum and classical correlations. Its definition is as follows [18],

$$I = S(\rho_A) + S(\rho_R) - S(\rho_{AR}) \qquad (6)$$

where $S(\rho) = -\text{Tr}(\rho \log_2 \rho) = -\sum_i \lambda_i \log_2 \lambda_i$ is von Neumann entropy for the quantum state $\rho$, and $\lambda_i$ is the eigenvalues of the state density operator $\rho$. $\rho_A$ and $\rho_R$ represent the reduced density matrix of the Alice part and the Rob part in the system respectively, $\rho_{A/R} = Tr_{R/A}(\rho_{AR})$.

Consider $\rho_{AR} = \rho'$ in Eq. (4), and after calculation, the von Neumann entropies are given as

$$S(\rho_A) = 1 \qquad (7)$$

$$S(\rho_R)=-\sum_{n_R=0}^{\infty}\frac{\tanh^{2n_R} r}{2\cosh^2 r}\left(1+\frac{n_R}{\sinh^2 r}\right)\log_2\left[\frac{\tanh^{2n_R} r}{2\cosh^2 r}\left(1+\frac{n_R}{\sinh^2 r}\right)\right] \quad (8)$$

$$S(\rho_{AR})=-\sum_{n_R}\frac{\tanh^{2n_R} r}{2\cosh^2 r}\left(1+\frac{n_R+1}{\cosh^2 r}\right)\log_2\left[\frac{\tanh^{2n_R} r}{2\cosh^2 r}\left(1+\frac{n_R+1}{\cosh^2 r}\right)\right] \quad (9)$$

where $n_R$ is the number of particles seen by the accelerated observer Rob.

The variation trend of the calculated mutual information with acceleration is shown in Fig. 2. As the acceleration increases, the whole correlation of quantum states gradually decreases and finally approaches to 1.

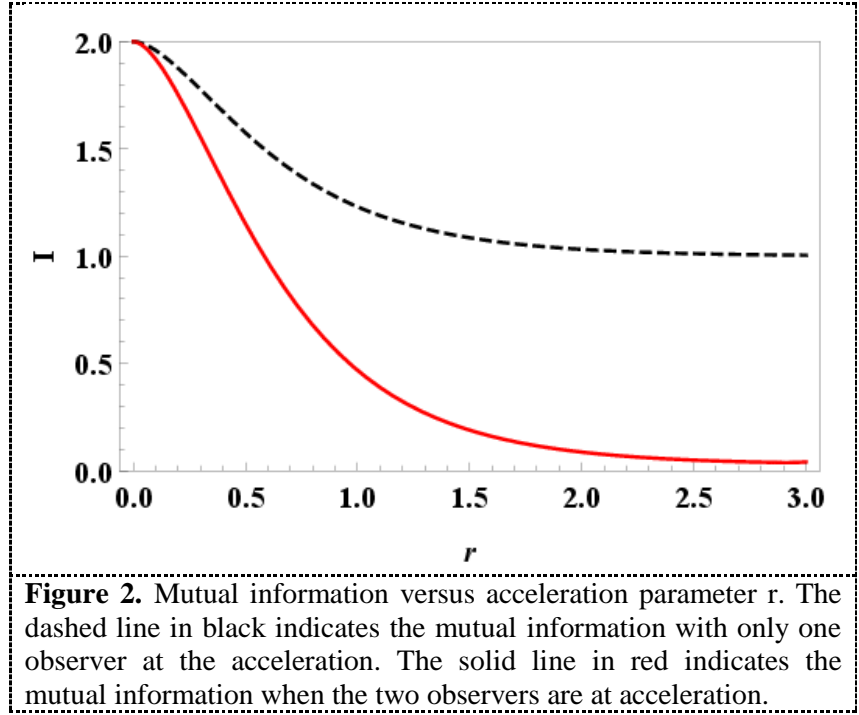

**Figure 2.** Mutual information versus acceleration parameter r. The dashed line in black indicates the mutual information with only one observer at the acceleration. The solid line in red indicates the mutual information when the two observers are at acceleration.

It is noted that the quantum correlation described by quantum discord had been investigated [17] when Rob accelerates, and it was found that although the entanglement would disappear at the finite acceleration, the quantum correlation approaches to a finite value. It means that the acceleration cannot break all the quantum correlation. In this section, we hope to investigate further the change of the classical correlation under the background of the Unruh effect as a noise quantum channel.

Here, the concept of the correlation is derived from the quantum discord in which the whole correlation represented by the mutual information could be divided into a quantum and a classical part [14,15]. This requires the conditional entropy,

$$S(R|A)=\sum_i p_i S(\rho_{R|i}) \quad (10)$$

which is the conditional entropy of $R$ part given the $i$ th measurement result of $A$.

The classical correlation can be calculated when the conditional entropy obtained by the selected measurement method is the maximum,

$$J=S(\rho_R)-S_{\max}(R|A) \quad (11)$$

In this paper, we consider the projection measurement which can reach at our purpose [17]. The projection measurement can be described as $[E_i]_{kl} = \delta_{ki}\delta_{li}$ where $i,k,l = 1,2$. The postmeasurement states can be written as

$$\rho_{R|0} = Tr_A(E_0 \rho_{AR} E_0)/p_0 = \sum_{n_R=0}^{\infty} \frac{\tanh^{2n_R} r_R}{\cosh^2 r_R} |n_R\rangle\langle n_R|$$

(12)

$$\rho_{R|1} = Tr_A(E_1 \rho_{AR} E_1)/p_1 = \sum_{n_R=0}^{\infty} \frac{\tanh^{2n_R} r_R}{\cosh^2 r_R} \frac{n_R+1}{\cosh^2 r} |n_R+1\rangle\langle n_R+1|$$

(13)

This indicates that the measurement is made by Alice, and $\rho_{R|0,1}$ represents the state of Rob part for the corresponding measurement result 0 or 1 obtained by Alice. The probability for the measurement result 0 or 1 can be calculated as $p_{0,1} = Tr_{AR}(E_{0,1} \rho_{AR} E_{0,1}) = \frac{1}{2}$. Actually, the same result would be obtained when the projection measurement is made by Rob.

Using Eqs. (12), (13) and (11), we can obtain the expression of classical correlation and its change with the acceleration is presented in Fig. 3. It is seen that the classical correlation would approach to a finite value when the acceleration increases.

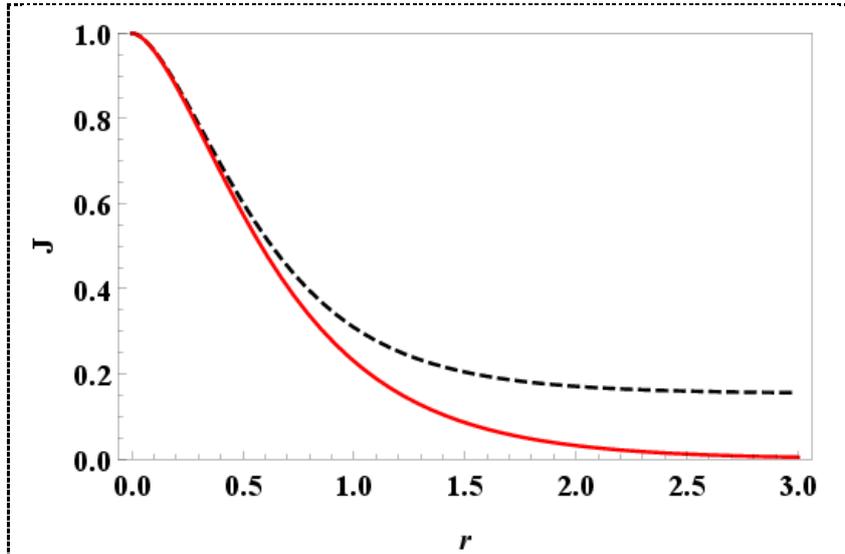

**Figure 3.** Classical correlation versus acceleration parameter r. The dashed line in black indicates the classical correlation with only one observer at the acceleration. The solid line in red indicates the classical correlation when the two observers are at acceleration.

For comparison, we also plot the quantum discord, $D = I - J$, as presented in Fig. 4. It is noted that the quantum correlation decreases not so much as that for the classical correlation in Fig.3 and the entanglement fidelity in Fig.1.

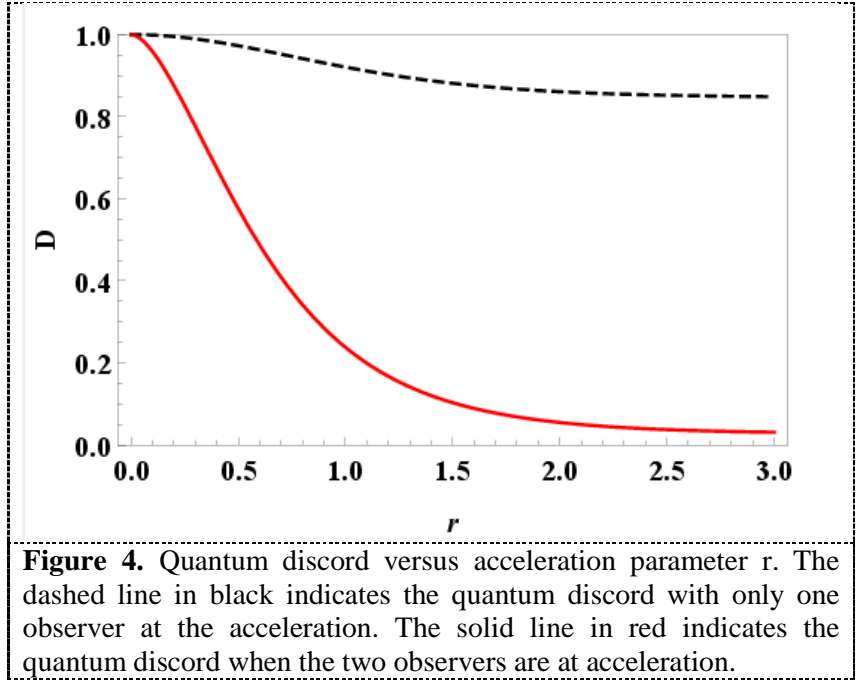

**Figure 4.** Quantum discord versus acceleration parameter r. The dashed line in black indicates the quantum discord with only one observer at the acceleration. The solid line in red indicates the quantum discord when the two observers are at acceleration.

## 4. Two observers in simultaneously acceleration

In this section, we continue to consider the case that both Alice and Rob have the acceleration, whose acceleration parameters are $r_A$ and $r_R$ respectively. In this case, the quantum state after the acceleration is calculated as

$$\rho_{AR} = \sum_{n_A=0}^{\infty}\sum_{n_R=0}^{\infty} 2C_{n_A}C_{n_R} \left[\begin{array}{l}|n_A n_R\rangle\langle n_A n_R| + \dfrac{n_A+1}{\cosh^2 r_A}\dfrac{n_R+1}{\cosh^2 r_R}|(n_A+1)(n_R+1)\rangle\langle(n_A+1)(n_R+1)| \\ + \dfrac{\sqrt{n_A+1}}{\cosh r_A}\dfrac{\sqrt{n_R+1}}{\cosh r_R}\left(|n_A n_R\rangle\langle(n_A+1)(n_R+1)| + |(n_A+1)(n_R+1)\rangle\langle n_A n_R|\right)\end{array}\right] \quad (14)$$

where $C_{n_A} = \dfrac{\tanh^{2n_A} r_A}{2\cosh^2 r_A}$ and $C_{n_R} = \dfrac{\tanh^{2n_R} r_R}{2\cosh^2 r_R}$.

By comparing the density operators of the states before and after the acceleration, it is found that the Unruh effect can still be considered as a quantum noise channel,

$$\rho \to \rho_{AR} = \sum_{n_A, n_R}^{\infty} A_{n_A, n_R} \rho A^{\dagger}_{n_A, n_R} \quad (15)$$

where the positive and trace preserving super-operator is obtained as

$$A_{n_A, n_R} = \dfrac{1}{\sqrt{n_A!}}\dfrac{1}{\sqrt{n_R!}} \tanh^{n_A} r_A \tanh^{n_R} r_R \left(b_A^{\dagger}\right)^{n_A} (\cosh r_R)^{-(\hat{n}_A+1)} \otimes \left(b_R^{\dagger}\right)^{n_R} (\cosh r_A)^{-(\hat{n}_R+1)} \quad (16)$$

Analyzing the entanglement fidelity, we get

$$Fe = \sum_n \left(Tr\rho_{AR} A_{n_A, n_R}\right)\left(Tr\rho_{AR} A^{\dagger}_{n_A, n_R}\right) = \dfrac{1}{4}\dfrac{1}{\cosh^2 r_A}\dfrac{1}{\cosh^2 r_R}\left(\dfrac{1}{\cosh r_R}+\dfrac{1}{\cosh r_A}\right)^2 \quad (17)$$

When $r_A = r_R = r$, the fidelity becomes

$$Fe = \frac{1}{4} \frac{1}{\cosh^4 r} \left(1 + \frac{1}{\cosh^2 r}\right)^2 \tag{18}$$

which is presented by the red solid line in Fig.1.

In order to understand the influence of Unruh effect on the quantum state (2) when Alice and Rob are accelerated simultaneously with the same value, we calculate the mutual information

$$I = S(\rho_A) + S(\rho_R) - S(\rho_{AR}) \tag{19}$$

where

$$S(\rho_{AR}) = -\sum_{n_A=0}^{\infty} \sum_{n_R=0}^{\infty} 2C_{n_A} C_{n_R} \left(1 + \frac{n_A+1}{\cosh^2 r} \frac{n_R+1}{\cosh^2 r}\right) \log_2 \left[2C_{n_A} C_{n_R} \left(1 + \frac{n_A+1}{\cosh^2 r} \frac{n_R+1}{\cosh^2 r}\right)\right] \tag{20}$$

$$S(\rho_R) = -\sum_{n_R=0}^{\infty} C_{n_R} \left(1 + \frac{n_R}{\sinh^2 r}\right) \log_2 \left[C_{n_R} \left(1 + \frac{n_R}{\sinh^2 r}\right)\right] \tag{21}$$

$$S(\rho_A) = -\sum_{n_A=0}^{\infty} C_{n_A} \left(1 + \frac{n_A}{\sinh^2 r}\right) \log_2 \left[C_{n_A} \left(1 + \frac{n_A}{\sinh^2 r}\right)\right] \tag{22}$$

The mutual information is presented in Fig.2. Different from the case that only one observer is accelerated, the mutual information for the quantum state influenced by two observers accelerating simultaneously would decrease to zero quickly with the acceleration.

Since mutual information is only a general description of the whole correlation, we carefully analyze the classical correlation and quantum correlation. Projection measurement is still used to calculate conditional entropy when calculating classical correlation. Projection measurement is performed on the part of Alice, and the measurement operator of its projection measurement is expressed as $[E_i]_{kl} = \delta_{ki}\delta_{li}$, where $i,k,l = 1,\cdots,\infty$.

After the measurement, we can get $\rho_{AR|n_A} = E_{n_A} \rho_{AR} E_{n_A}$, $\rho_{AR|n_A+1} = E_{n_A+1} \rho_{AR} E_{n_A+1}$. The probability of measurement can be calculated by tracing out the Alice part and the Rob part, $p_{n_A} = Tr_{AR}(E_{n_A} \rho_{AR} E_{n_A})$, $p_{n_A+1} = Tr_{AR}(E_{n_A+1} \rho_{AR} E_{n_A+1})$. In order to get the classical correlation $J = S(\rho_R) - S(R|A)$, we have to calculate the conditional entropy $S(R|A) = \sum_i p_i S(\rho_{R|i})$ where $\rho_{R|i} = Tr_A(E_i \rho_{AR} E_i) / p_i$ acquired by

$$\rho_{R|n_A} = \sum_{n_R=0}^{\infty} \frac{\tanh^{2n_R} r}{\cosh^2 r} |n_R\rangle\langle n_R| \tag{23}$$

$$\rho_{R|n_A+1} = \sum_{n_R=0}^{\infty} \frac{\tanh^{2n_R} r}{\cosh^2 r} \frac{n_R+1}{\cosh^2 r} |(n_R+1)\rangle\langle(n_R+1)| \tag{24}$$

Then the classical correlation can be calculated according to Eq. (11), but the entropies is related to the case for two observers accelerating simultaneously. See Fig. 3 for the presentation of classical correlation, and it is found that the classical correlation would decrease to zero with the increase of the acceleration.

After the classical correlation has been obtained, the physical representation of the quantum correlation can be obtained by subtracting the classical correlation from the mutual information, $D = I - J$. Its change with acceleration is shown by the red solid line in Fig. 4.

## 5. Conclusion

The Unruh effect can be regarded as the noise channel, with that the entanglement fidelity of any initial entangled state would be reduced to zero with the increase of acceleration for only one or two observers. But, when both observers have accelerations, the entanglement fidelity decreases more obviously. We have also studied the change of the correlation for the quantum entangled state influenced by the acceleration for only one or two observers. It is interesting that the quantum and classical correlation would approach to a finite value when only one observer is accelerated, but they would decrease to zero quickly with the increase of the acceleration when two observers are accelerated at the same value. As well-realized, when the acceleration is caused by the gravitational field, it implies that the quantum or the classical correlation is not easy to be preserved for the observers stayed in the gravitational field. This conclusion is interesting and significant for understanding the curved space-time and quantum entanglement.